\documentclass[journal]{IEEEtran}

\ifCLASSINFOpdf
   \usepackage[pdftex]{graphicx}
   \graphicspath{{../figures/}}
   \DeclareGraphicsExtensions{.pdf,.jpeg,.png}
\else
   \usepackage[dvips]{graphicx}
   \graphicspath{{../figures/}}
   \DeclareGraphicsExtensions{.eps}
\fi
\ifCLASSOPTIONcompsoc
  \usepackage[caption=false,font=normalsize,labelfont=sf,textfont=sf]{subfig}
\else
  \usepackage[caption=false,font=footnotesize]{subfig}
\fi
\usepackage{url}

\usepackage{xcolor}
\usepackage[center]{caption}

\usepackage{tikz}

\newcommand\copyrighttext{%
  \footnotesize This work has been submitted to the IEEE for possible publication. Copyright may be transferred without notice, after which this version may no longer be accessible}
\newcommand\copyrightnotice{%
\begin{tikzpicture}[remember picture,overlay]
\node[anchor=south,yshift=10pt] at (current page.south) {\fbox{\parbox{\dimexpr\textwidth-\fboxsep-\fboxrule\relax}{\copyrighttext}}};
\end{tikzpicture}%
}

\begin{document}

%
\title{MusicID: A Brainwave-based User Authentication System for Internet of Things}

\author{
		Jinani Sooriyaarachchi,
		Suranga Seneviratne,
		Kanchana Thilakarathna, and
		Albert Y. Zomaya
\thanks{ J. Sooriyaarachchi was with Data61-CSIRO, Australia.}
\thanks{ S. Seneviratne, K. Thilakarathna and A. Y. Zomaya are with School of Computer Science, The University Sydney, Australia.}}

\maketitle

\begin{abstract}
We propose MusicID, an authentication solution for smart devices that uses music-induced brainwave patterns as a behavioral biometric modality. We experimentally evaluate MusicID using data collected from real users whilst they are listening to two forms of music; \emph{a popular English song} and \emph{individual's favorite song}. We show that an accuracy over 98\% for \emph{user identification} and an accuracy over 97\% for \emph{user verification} can be achieved by using data collected from a 4-electrode commodity brainwave headset. We further show that a single electrode is able to provide an accuracy of approximately 85\% and the use of two electrodes provides an accuracy of approximately 95\%. As already shown by commodity brain-sensing headsets for meditation applications, we believe including dry EEG electrodes in smart-headsets is feasible and MusicID has the potential of providing an \emph{entry point} and \emph{continuous} authentication framework for upcoming surge of smart-devices mainly driven by Augmented Reality (AR)/Virtual Reality (VR) applications.
\end{abstract}

\copyrightnotice

\section{Introduction}
\label{Sec:Introduction}

\IEEEPARstart{S}{mart} head-mounted devices of various kinds such as smart glasses, 
mixed reality headsets, and  smart earbuds, are becoming increasingly popular in engaging and controlling smart Internet of Things (IoT) environments. This is in addition to the commonly used Bluetooth headsets which are the most sold wearable globally by a significant margin~\cite{gartner2017}. Moreover, the rapid growth of mixed reality in multiple application domains such as gaming, education, and health will also ensure that the usage of smart-headsets, especially the ones that are capable of standalone operation will continue to increase. Secure and robust user authentication on such headsets is of paramount importance due to unprecedented sensitivity of information that is sensed, stored, and transmitted by these devices such as health-care data, financial records, user location, and surrounding environmental information. In industrial and health-care settings, it is necessary to make sure that these devices are worn by the authorized personnel only. Also, in personal smart-home settings, authentication of the user enables personalized service delivery.

Existing methods of authentication such as \emph{passwords and PINs} are not observation resistant and less secure due to reuse. Attempts to increase the complexity of passwords by adding more constraints on character, numeral, and symbol combinations, reduce the usability. Other alternative of \emph{static biometrics} such as fingerprints and face IDs are prone to replay attacks and spoofing~\cite{matsumoto2002impact}. As a result, \emph{behavioral biometrics} are emerging as an exciting new alternative. Behavioral biometrics focus on behavior modalities that are potentially unique to individuals in the likes of typing patterns~\cite{monrose2000keystroke}, touch patterns~\cite{frank2013touchalytics}, 
gait~\cite{gafurov2006biometric}, and heart rate~\cite{article}.
Behavioral biometrics are more secure as they are difficult to mimic, record, and synthesize and can readily be used for implicit authentication~\cite{saeed2016new}. The main challenge in  behavioral biometrics is finding modalities that can be captured non-intrusively and are sufficiently consistent over time so that they can be used for user re-identification. Incorporating behavioral biometrics to standalone Internet of Things devices such as smart-headsets is further challenging due to their limited user interaction capabilities.



In this paper, we propose \emph{MusicID}, a novel behavioral biometric for user authentication for Internet of Things environments, which measures the emotional behavior and reactions of the user to music through EEG ({\bf E}lectro{\bf e}ncephalo{\bf g}ram) signals. Recent advances in head-mounted wearables made non-invasive EEG capture possible through soft dry electrodes as demonstrated by commodity headsets such as Muse~\cite{muse2017} and Neurosky~\cite{neurosky2017}. These sensors can be easily integrated with smart-headsets that are increasingly becoming popular in both smart-home and industrial Internet of Things.
The possibility of using brainwave patterns for authentication makes inroads towards\ a unique and hard to spoof behavioral biometric modality that can be useful not only for smart-headsets such as Microsoft Holo Lens and Oculus Rift, but also in general as a universal authentication solution.


We make following contributions in this paper.

\begin{itemize}
\item We show that the music stimulated EEG patterns are potentially unique to individuals by collecting EEG data from 20 users over one and half months using Muse brain sensing headband. We measure brain reactions to each individuals' \emph{favorite song} as well as a \emph{reference song} and show that our observations are statistically significant through an ANOVA test ($p<0.0001$). 

\item We develop Random Forest classifiers for both \emph{user identification} and \emph{user verification} tasks using features extracted from decomposed EEG signals and show that over 97\% accuracy can be achieved.


\item We investigate the trade-off between accuracy and the number of electrodes and frequency bands. Our results show that use of two front electrodes gives an accuracy of 94\%. Also, we show that all four brainwave frequency bands are necessary to have a significant accuracy.

\end{itemize}


The rest of the paper is organized as follows. In Section \ref{Sec:Background}, we provide background information on EEG and brain behaviors. Section \ref{Sec:Methodology} describes our methodology and Section \ref{Sec:Results} presents results. We discuss related work in Section \ref{Sec:RelatedWork} followed by a discussion of implications, limitations, and future work in Section \ref{Sec:Discussion}. Section \ref{Sec:Conclusion} concludes the paper.

\section{Background}
\label{Sec:Background}

EEG ({\bf E}lectro{\bf e}ncephalo{\bf g}ram) measures the electrical activities in human brain. EEG data is captured by measuring voltage difference between two electrodes placed on the scalp. More specifically EEG data consist of inhibitory and excitatory post-synaptic potentials generated in pyramidal cells of the brain cortex depending on the mental activity. Whenever the brain receives stimuli from our senses such as visual, audio, touch, pain, taste, and smell, EEG signals vary in magnitude as well as in frequency. EEG brain signals can be decomposed into five based on their frequencies; \emph{Alpha}, \emph{Beta}, \emph{Theta}, \emph{Delta}, and \emph{Gamma}. Each of these frequency bands has its own characteristics depending on the brain activity and the state of consciousness. Table~\ref{Tab:Waves} summarizes the frequencies of each brainwave band and activities that generate them. 

 \begin{table}[!t]

\scriptsize
\caption{Summary of different brainwaves} \vspace{-1mm}
\label{Tab:Waves}
\centering
\begin{tabular}{p{0.7cm}p{1.45cm}p{4.8cm}}
{\bf Type } & {\bf Frequency} & {\bf Activities} \\ \hline

Delta & 0.5-4.0 Hz& Deep sleep\\ \hline
Theta & 4.0-7.5 Hz& Meditation, Light sleep, Vivid visualizations, Creativity, Emotional connection, Relaxation\\ \hline
Alpha & 7.5-12.0 Hz& Deep relaxation with eyes closed, State of concentration, Memorizing, Learning, Visualization\\ \hline
Beta & 12.0-30.0 Hz& Normal waking conditions, Daily consciousness \\ \hline
Gamma & 30.0-100.0 Hz& High level information processing in improved memory conditions, Sudden insights\\ \hline

\end{tabular} \vspace{-2mm}

\end{table}



Brain structure of every human being is unique and highly influenced by DNA~\cite{thompson2001genetic}. Also, how a person's brain responds to a stimulus is highly dependent on that person's previous experiences. As such,  brainwave patterns generated for some activities can potentially be used as an authentication modality as they can be unique to individuals and difficult to spoof.

\begin{figure}[!t]
\centering
\includegraphics[trim=0cm 0cm 0cm 0cm, clip=true,scale=0.33]{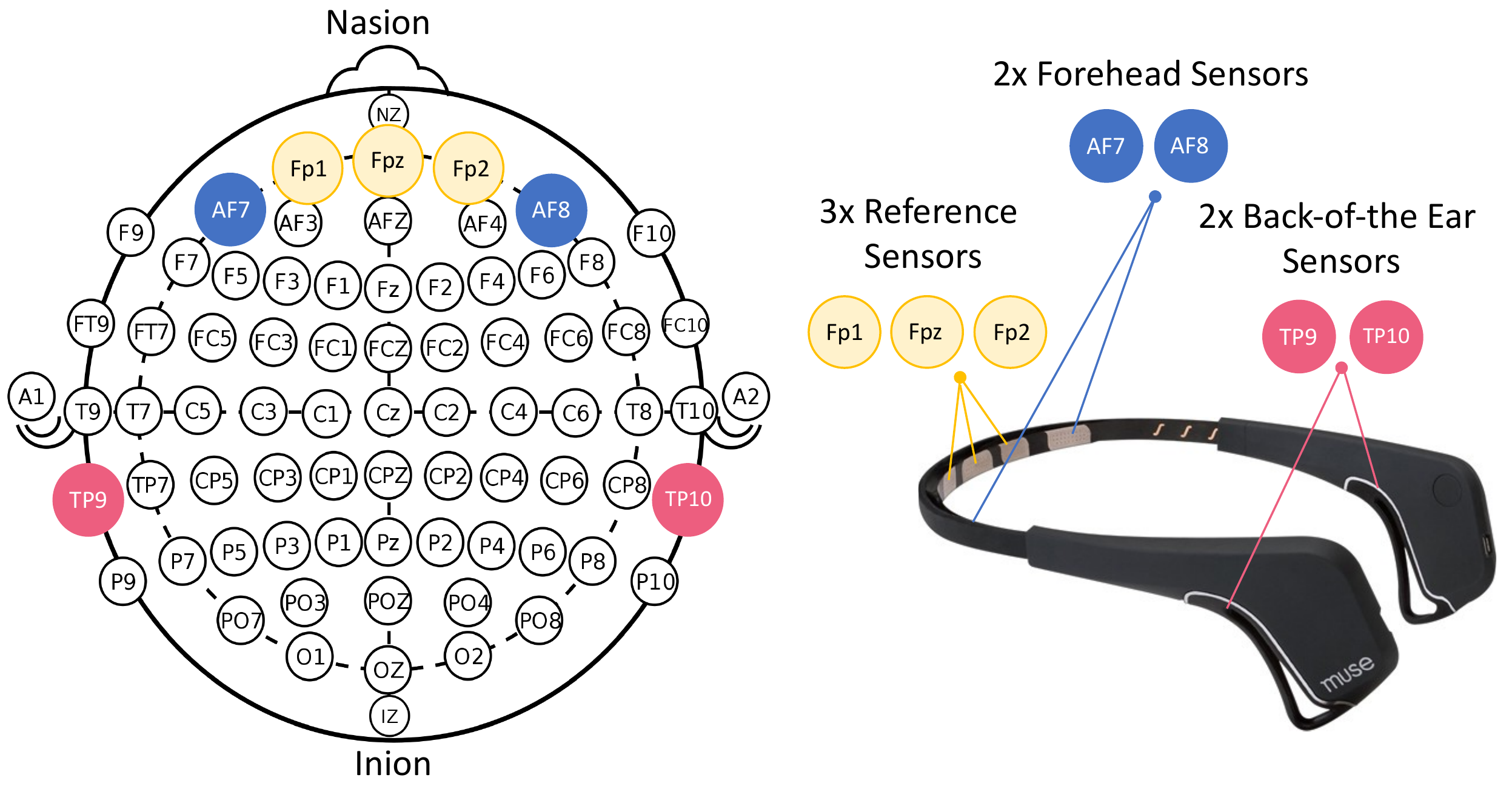}
\caption{Standard 10-20 system} \vspace{-2mm}
\label{Fig:10-20}
\end{figure}

Brainwaves can be measured using \emph{invasive techniques} in which electrodes are placed under the scalp or \emph{non-invasive techniques} in which electrodes are placed above the scalp. Scalp electrodes can detect faint electrical signals on the surface of the head that represent the underlying neural activity. Figure~\ref{Fig:10-20} shows the standard 10-20  electrode locations commonly used to place electrodes on the scalp.



Various studies found changes in brainwaves induced by the emotions and aesthetic pleasure of music.
These include \emph{Beta} bands playing a major role in music processing~\cite{10.2307/40285613},\emph{Gamma} brainwaves confined to subjects with music training~\cite{bhattacharya2001musicians}, and high \emph{Theta} brainwave power in frontal midline in contrast to pleasant and unpleasant music~\cite{sammler2007music}. Motivated by these background information, in this paper, we explore the feasibility of using music-induced brainwaves as a behavioral biometric modality.


\section{Methodology}
\label{Sec:Methodology}

\begin{figure}[!t]
\centering
\includegraphics[trim=0cm 0cm 0cm 0cm, clip=true,scale=0.38]{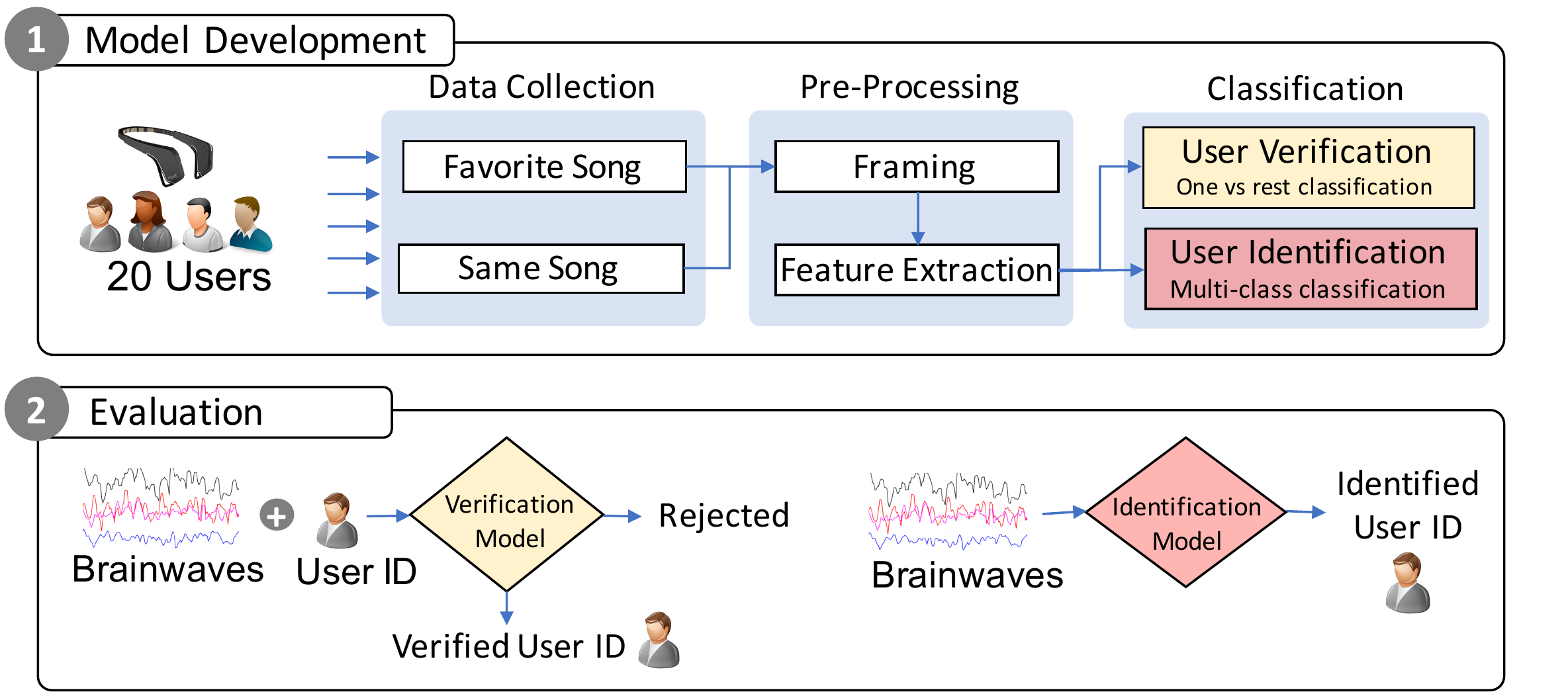}
\caption {Methodology} \vspace{-4mm}
\label{Fig:Flow}
\end{figure}

In this section we describe our experiment design and data collection process followed by data pre-processing steps, features used, and classifier details. The overall summary of our methodology is illustrated in Figure~\ref{Fig:Flow}.

\subsection{Data Collection}

We recruited 20 volunteers (11 females and 9 males, aged between 20 and 40) to participate in our experiment. The participants were students and research staff associated with Data61, CSIRO. The duration of the experiment spanned over two months and there was a minimum one week gap between two sessions of a single participant. Each user participated in at least two sessions. Our experiment was approved by the CSIRO Human Research Ethics Committee (Ref. No. 104/17). 

During each session the participants performed two  tasks; \emph{i) listening to a popular English song\footnote{\url{https://www.youtube.com/watch?v=JRfuAukYTKg}}} and \emph{ii) listening to individual's favorite song} whilst wearing Muse~\cite{muse2017} brain sensing headset. Muse headset is equipped with 4 electrodes in the standard 4-channel configuration (TP9, AF7, AF8, TP10 as shown in Figure \ref{Fig:10-20}).  The duration of the each experiment was 150 seconds and the participants closed their eyes during this experiment. This was done due to two main reasons; \emph{i) avoid distractions and head movements that may affect the measurements}, and ii) \emph{eye blinks are known to impact the EEG readings}~\cite{21b1fabdb2fd41bfbfc93bcf6c5ba59a} as a result of the large potentials evoked above and below the eye during eye blinks and eye movements. 



\begin{figure}[h]
\centering
\includegraphics[trim=0cm 0cm 0cm 0cm, clip=true,scale=0.3]{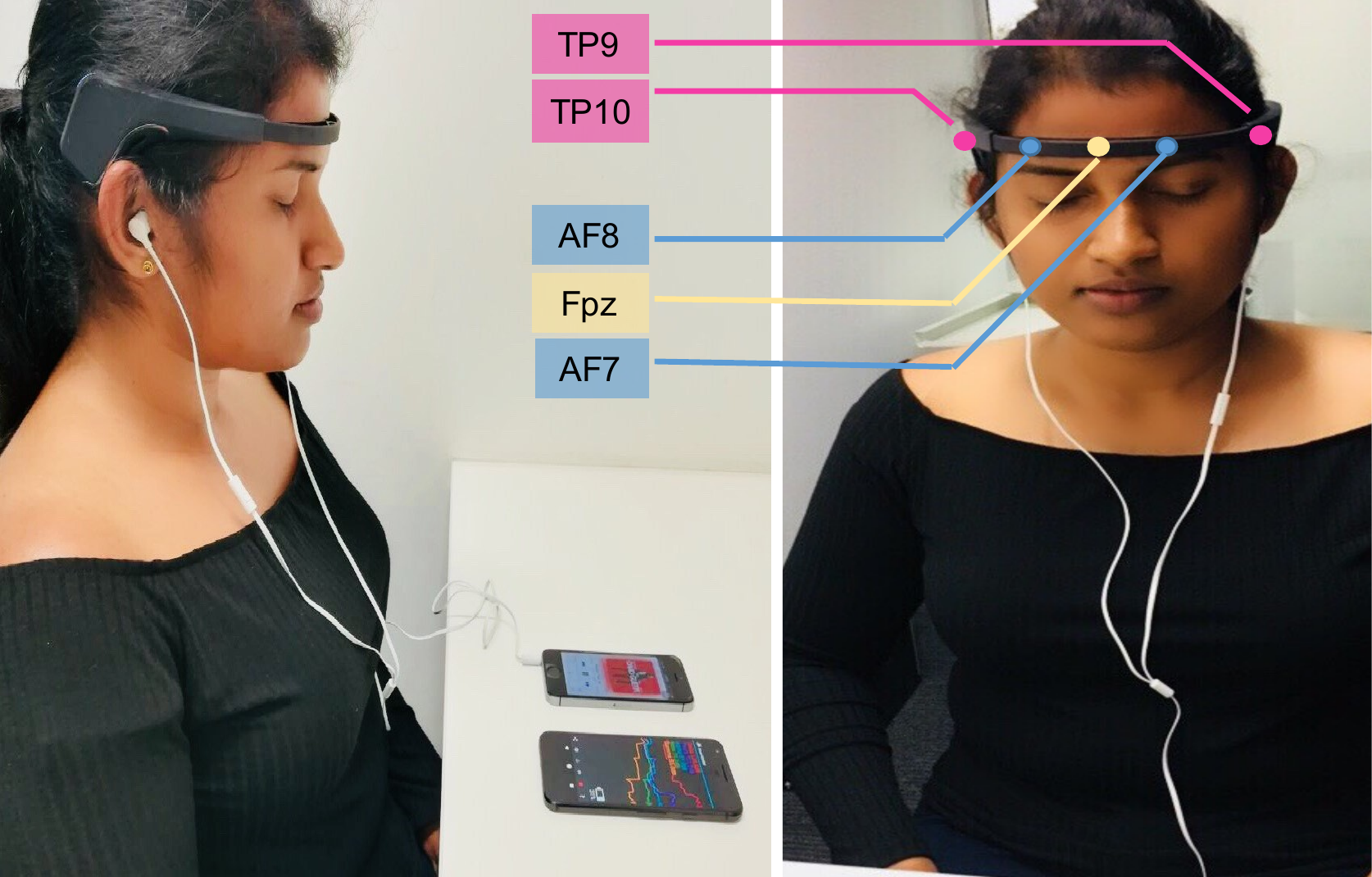}
\caption{Experimental setup} \vspace{-5mm}
\label{Fig:Exper}
\end{figure}
\begin{figure}[h]
\centering
\subfloat[Listening to Favorite Song]{\label{Fig:favsong}\includegraphics[trim=3cm 10cm 4.5cm 10.8cm, clip=true,scale=0.5]{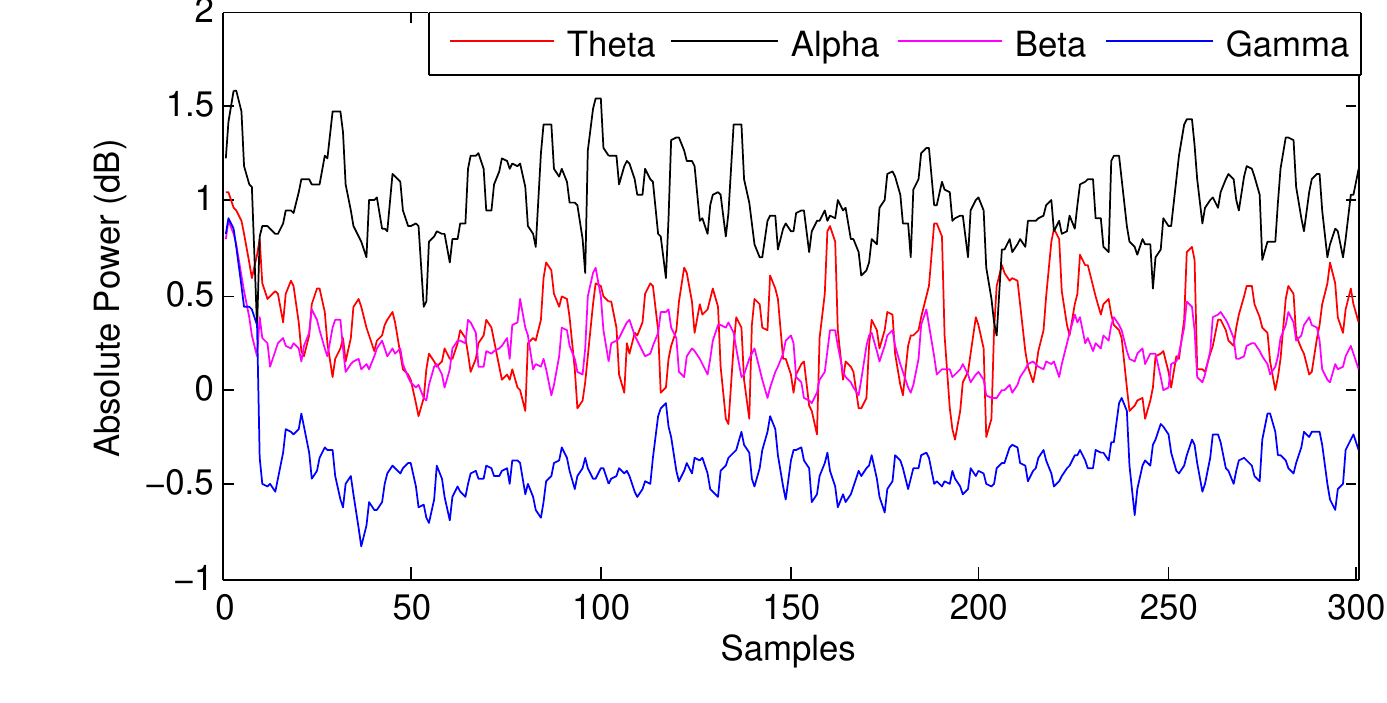}}\\\vspace{-2mm}
\subfloat[Listening to Same Song]{\label{Fig:same}\includegraphics[trim=3cm 10cm 4.5cm 10.8cm, clip=true,scale=0.5]{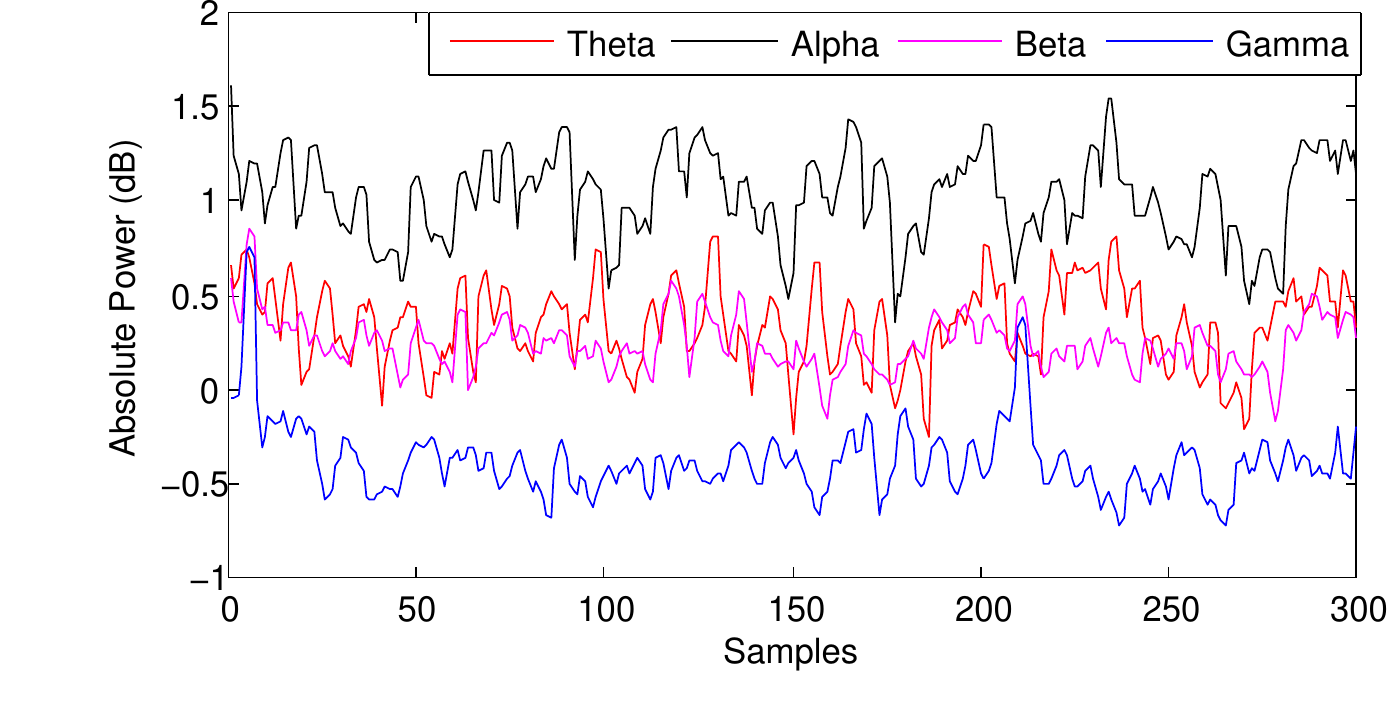}}\vspace{-2mm}
\caption{Example brainwave data streams from AF7} \vspace{-3mm}
\label{Fig:waves}
\end{figure}

In the Muse Monitor application~\cite{musemonitor2017} running in a Google Pixel phone, we selected a \emph{sampling rate} of 220 Hz and  a \emph{recording interval} of 0.5 seconds. The recording interval defines when the data is recorded to the output $.csv$ file. Thus, at the end of each session with this recording interval, 300 data samples were collected for the task of listening to the reference song and another 300 data samples for the task of listening to favorite songs. Each sample included 24 readings; absolute brainwave values of \emph{Alpha}, \emph{Beta}, \emph{Theta}, \emph{Delta}, \emph{Gamma}, and \emph{raw EEG} (without separating into the sub bands) from each of 4-channels. Figure~\ref{Fig:Exper} shows the setup of the experiment and in Figure~\ref{Fig:waves}  we show example brainwave data streams of a user collected from AF7 electrode for two experiment scenarios.

\subsection{Data pre-processing and features}

Since \emph{Delta} brainwave frequencies are related to deep sleep, 
and not significant in our experimental tasks, we removed 4 absolute \emph{Delta} brainwave readings from 4-channels and used the remaining 20 readings.

\begin{figure}[h]
\centering
\includegraphics[trim=0cm 0cm 0cm 0cm, clip=true,scale=0.4]{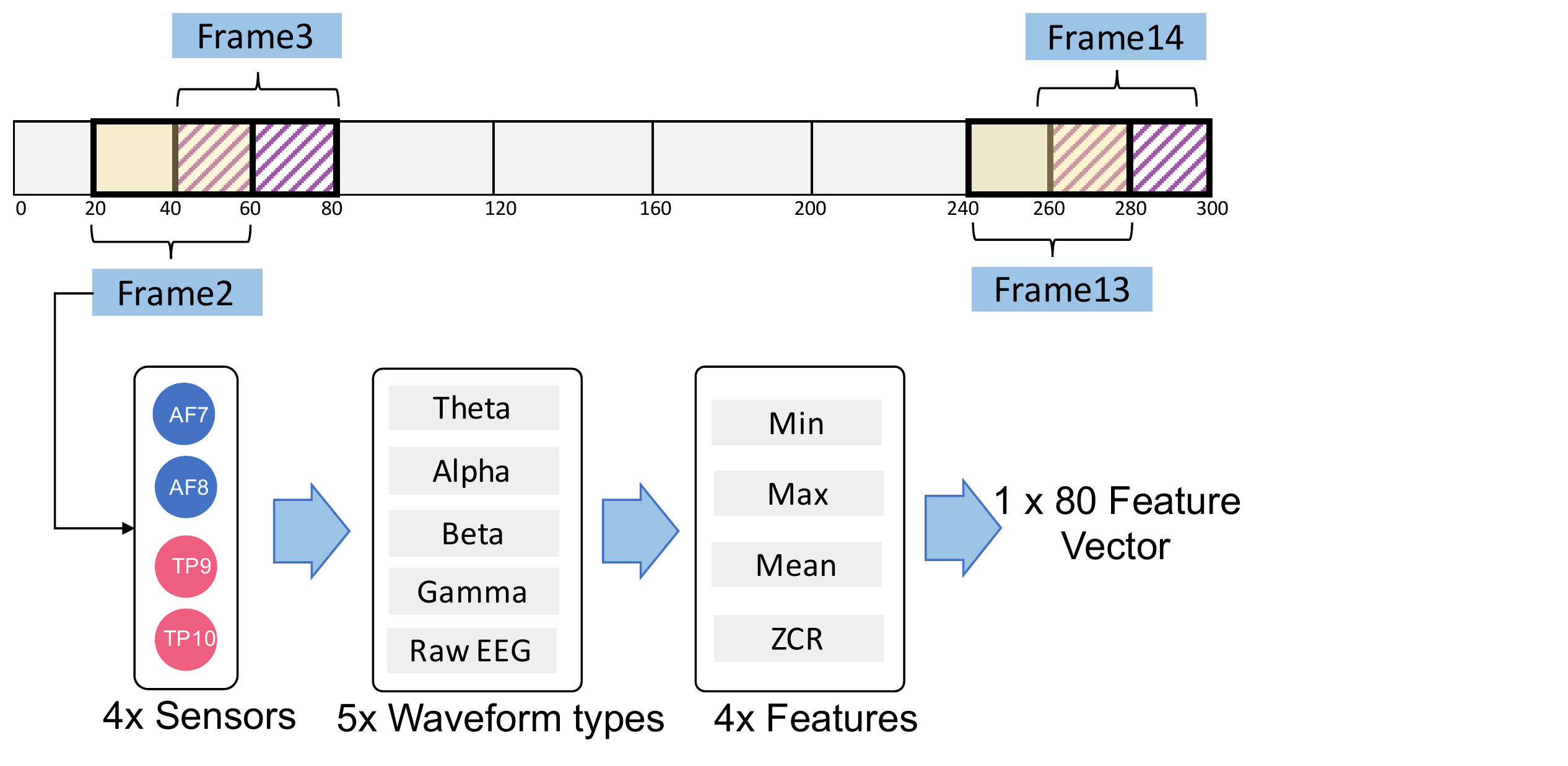}
\caption{Framing 300 samples to 14 frames} \vspace{-5mm}
\label{Fig:Ex4}
\end{figure}

We framed 300 samples from each session into 14 overlapping frames of 50\% overlap as shown in Figure~\ref{Fig:Ex4}. For each frame we had 20 readings for \emph{Alpha}, \emph{Beta}, \emph{Theta}, \emph{Gamma}, and \emph{raw EEG} values coming from each of the four electrodes. For each reading we calculated the \emph{mean}, \emph{maximum}, \emph{minimum}, and \emph{zero crossing rate (ZCR)} giving us 80 features for each frame. Thus, one listening task in a session gave us 14 samples of data that are of size $1 \times 80$. 




\subsection{Classifier}
\label{SubSec:Classifier}

We built Random Forest classifiers for two authentication settings; \emph{user identification} and \emph{user verification}. The choice of Random Forest is empirically decided based on the fact that it performed better compared to Support Vector Machines and Logistic Regression, which is evaluated in Section \ref{sec:classifier_results}.

For \emph{user identification}, we used Random Forest classifiers in a multi-class setting. For \emph{user verification} we used Random Forest classifiers in One-vs-Rest configuration where a classifier is built on per user basis that predicts whether a given frame belongs a particular user or not. To avoid over-fitting we performed 10-fold cross validation. We used 80\% of the samples from each session for training and the rest of the 20\% for testing. We avoided two consecutive feature vectors going into training and test sets since there is a 50\% overlap and allowing it would cause information transfer from the training set to the test set, artificially increasing the classifier performance.  Table~\ref{Tab:Samples} shows the number of samples for each user in training and test sets. This training and test dataset sizes are identical for both experiment scenarios.




\begin{table}[h]
\scriptsize
\centering
\caption{Summary of training and test datasets} \vspace{-1mm}
\begin{tabular}{p{1.3cm}ccc}
{\bf Users } & {\bf No. of sessions} & {\bf Training  set size} & {\bf Test  set size}\\ \hline
User 1-5 & 5 &55 & 15\\ \hline
User 6& 4 & 44  & 12\\ \hline
User 7-11& 3 &33 & 9\\ \hline
User 12-20& 2 &22 & 6\\ \hline

{\bf Total}& &{\bf 682} & {\bf 186}\\ \hline
\end{tabular} \vspace{-2mm}
\label{Tab:Samples}
\end{table}

%

\section{Results}
\label{Sec:Results}
First, we present an analysis of our features using hypothesis testing followed by the classifier performance for user identification and verification tasks. Then, we further analyze the data to identify the most important features that differentiate users. Next, we provide results to show how the placement of electrodes affect the accuracy and whether we can use only a certain type of brainwaves for authentication. Finally, we do cross song training and testing to check whether the user responds to music in general or there are differences in responses when listening to favorite music and other music.  

\subsection{Statistical hypothesis testing}

To check whether the features we extracted from brainwaves are specific to individual users, we conducted a \emph{statistical hypothesis testing} under the null hypothesis; \emph{``features extracted from brainwaves do not have a relationship to the corresponding user''} using ANOVA (Analysis of Variance). The resulting $F-test$ value was $22.462$ which corresponds to a $p$ value less than $0.0001$, indicating that null hypothesis can be omitted and the features indeed have a statistical relationship to the individual users.

\subsection{Classifier performance} 
\label{sec:classifier_results}
\noindent{{\bf \emph{i) User identification:}}} In user identification, given a brainwave sample the classifier predicts whose sample is that from a closed set of users. We trained Random Forest classifiers at different maximum depth levels using 10-fold cross validation for the two experiment scenarios. That is, we used the training dataset from listening to the same song and tested the models on the test set built using the same experiment scenario. Table~\ref{Tab:UserIdentification} shows the performance of the classifiers.

\begin{table}[h]
\scriptsize
\centering
\caption{User identification: accuracy on test set} \vspace{-1mm}
\begin{tabular}{p{1.5cm}|p{1.55cm}|p{1.55cm}}
{\bf Maximum Tree Depth } & {\bf Listening to same song} & {\bf Listening to favorite song}\\ \hline
5 & 86.56\% &86.02\% \\ \hline
10 & 98.39\% & 98.39\%  \\ \hline
15 &98.39\%  & 99.46\%\\ \hline

\end{tabular} \vspace{-2mm}
\label{Tab:UserIdentification}
\end{table}

We obtained an accuracy of 98.39\% for listening to the same song and an accuracy of 99.46\% for listening to the favorite song with Random Forest classifier with a maximum tree depth of 15. This corresponds only to 3 (1.61\%) and 1 (0.54\%) incorrectly classified samples in the test set. The accuracy does not increase further along with the tree depth.



\noindent{{\bf \emph{ii) User verification:}}} In user verification the classifier is given with a sample and an user ID and needs to make a decision whether the given sample belongs to the said user or not. We show the performance of the classifier in Table~\ref{Tab:UserVerification}. We were able to achieve an accuracy of 98.92\% for listening to the same song and an accuracy of 97.31\% for listening to the favorite song with a maximum tree depth of 10 for the Random Forest classifier in One-vs-Rest configuration. We highlight that the classifier performance does not increase after the maximum depth of 10 and it is always good to select a classifier with lower depth given the choice to have a better generalization.



\begin{table}[h]
\scriptsize
\centering
\caption{User verification: accuracy on test set} \vspace{-1mm}
\begin{tabular}{p{1.5cm}|p{1.55cm}|p{1.55cm}}
{\bf Maximum Tree Depth } & {\bf Listening to same song} & {\bf Listening to favorite song}\\ \hline
5 & 98.39\% & 96.77\%\\ \hline
10 & 98.92\%  & 97.31\%\\ \hline
15 & 98.92\% & 97.31\%\\ \hline

\end{tabular} \vspace{-2mm}
\label{Tab:UserVerification}
\end{table}

\begin{figure}[t]
\centering \vspace{-3mm}
\subfloat[User identification]{\label{Fig:ConfusionUserIdent}\includegraphics[trim=0cm 0cm 0cm 0cm, clip=true,scale=0.14]{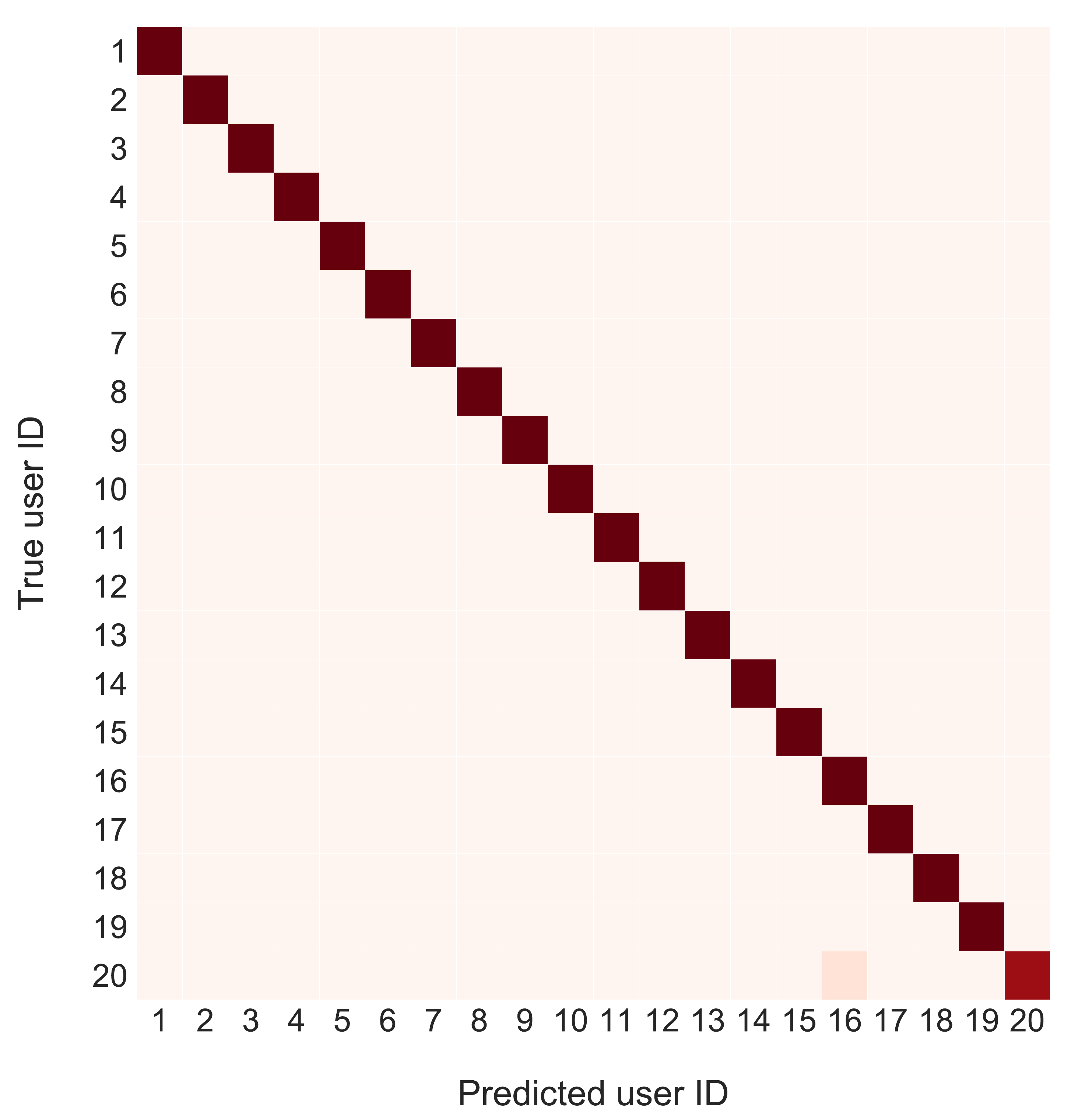}}
\subfloat[User verification]{\label{Fig:ConfusionUserVeri}\includegraphics[trim=0cm 0cm 0cm 0cm, clip=true,scale=0.14]{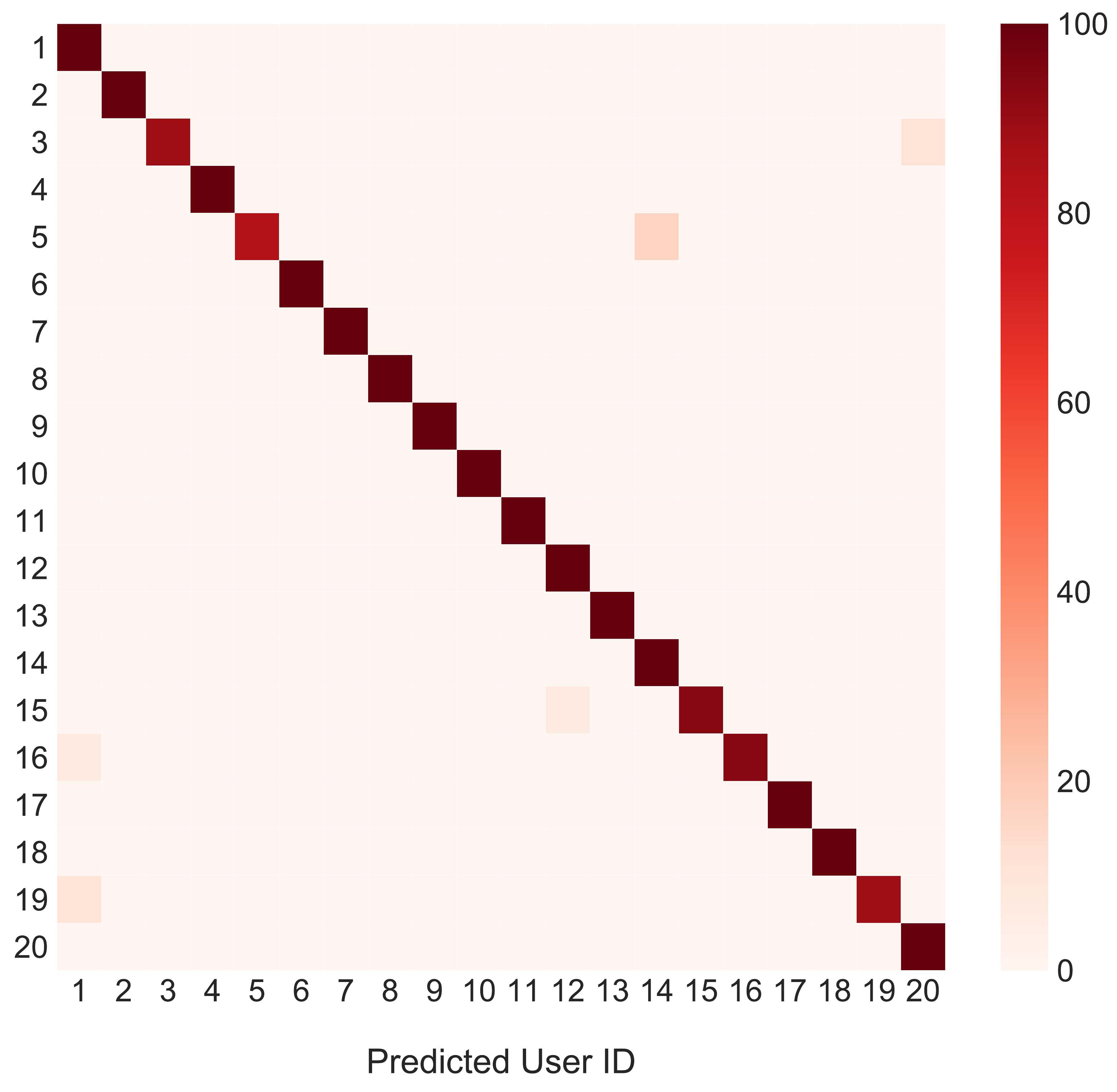}}
\caption{Confusion matrices of best classifiers (favorite song)}\vspace{-4mm}
\label{Fig:Confusion}
\end{figure}

In Figure~\ref{Fig:Confusion} we show some example confusion matrices. Figure~\ref{Fig:ConfusionUserIdent} shows the confusion matrix of the best classifier for user identification using the favorite song that resulted an accuracy of 99.46\%. The only error it makes is, it identifies one sample from \emph{User-20} as \emph{User-16}. Similarly, Figure~\ref{Fig:ConfusionUserVeri} show the best performing classifier for user verification using the favorite song that gave an accuracy of 97.31\%. 
Note that there is no significant difference in accuracy in the two experiment scenarios; listening to same song and listening one's favorite song. We analyze this further in Section~\ref{SubSec:Cross}.

Finally, as mentioned in Section~\ref{SubSec:Classifier} we selected Random Forest classifier because it was the best performing model out of the models we tested. For example, the SVM classifier achieved only 86.56\% and 82.26\% for user identification and verification tasks respectively whilst listening to the favorite song. For the task of listening to same song we were able to obtain an accuracy of 83.87\% for user identification and an accuracy of 86.55\% for user verification.

\subsection{Feature importance \& characteristics}
\label{SubSec:Importance}

To further understand the classifier performance we next do a feature importance analysis. We used the \emph{gini impurity} based feature importance generated using Random Forest classifiers.
\begin{figure}[h!]
\centering \vspace{-2mm}
\includegraphics[trim=0cm 0cm 0cm 0cm, clip=true,scale=0.45]{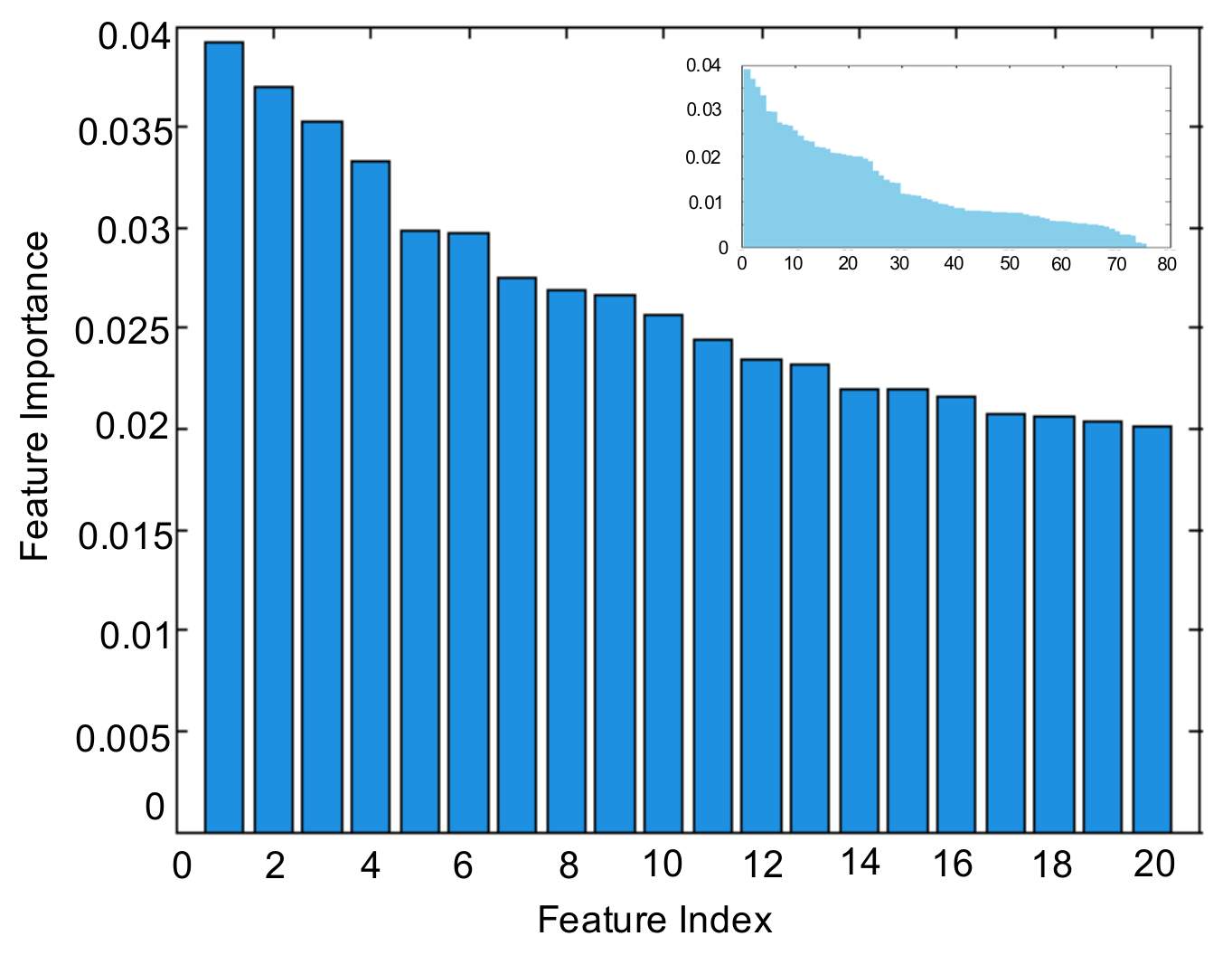}
\caption{Significance of individual features} \vspace{-2mm}
\label{Fig:Significance}
\end{figure}

Figure~\ref{Fig:Significance} shows the distribution of feature significance and Table~\ref{Tab:FeatureImportance} lists the top-20 features for the favorite song scenario. The same song scenario shows a similar behavior with only 2 changes in top 20 features. The mean values of \emph{Alpha} brainwaves obtained from all 4 electrodes show a high feature importance because the experiment subjects are closing their eyes and listening to the songs, making the \emph{Alpha} brainwave intensity higher as seen in Figure~\ref{Fig:waves}. Noticeably there are only three features coming from \emph{Gamma} waves which is potentially due to their low intensities as previously shown in Figure~\ref{Fig:waves}. 

\begin{figure}[h!]
\centering \vspace{-2mm}
\subfloat[Mean Alpha-TP9]{\label{Fig:MeanAlphaTP9}\includegraphics[trim=0cm 2cm 0cm 2cm, clip=true,scale=0.55]{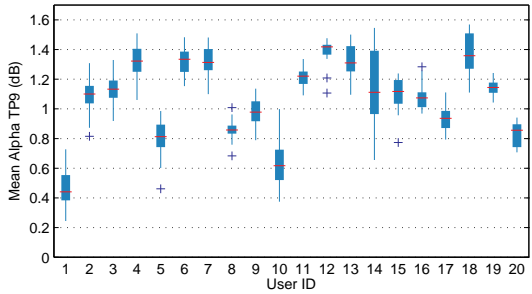}}\\
\subfloat[Mean Alpha-AF7]{\label{Fig:MeanAlphaAF7}\includegraphics[trim=0cm 2cm 0cm 2cm, clip=true,scale=0.55]{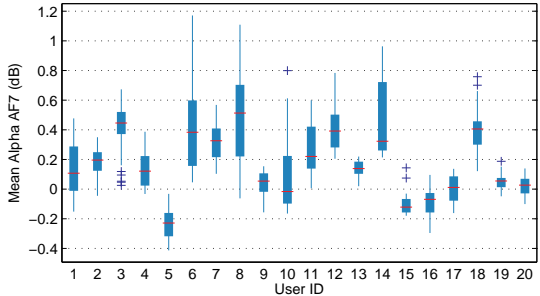}}
\caption{User-wise distribution of example significant features}
\label{Fig:ImFeatures} \vspace{-4mm}
\end{figure}

We show box plots of two example features of highest importance; \emph{mean of the Alpha waves measured at electrode position TP9} in Figure~\ref{Fig:MeanAlphaTP9} and \emph{mean of the Alpha waves measured at electrode position AF7} in Figure~\ref{Fig:MeanAlphaAF7}. As Figure~\ref{Fig:MeanAlphaTP9} shows, for many of the users the signal strength of \emph{Alpha} waves vary in a limited range in TP9 electrode with the exception of \emph{User-14}, \emph{User-18} and \emph{User-10}. Figure~\ref{Fig:MeanAlphaAF7} shows that in AF7 electrode the \emph{Alpha} wave signal strength varies more in number of users such as \emph{User-6}, \emph{User-8}, and \emph{User-14} compared to TP9 electrode. This is potentially due to the variation in regional activity of the brain stimulated by music and the fact that TP9 electrode is relatively closer to the auditory cortex than AF7 electrode. 

\begin{figure}[h!]
\centering
\includegraphics[trim=0cm 0.65cm 0cm 0.65cm, clip=true,scale=0.55]{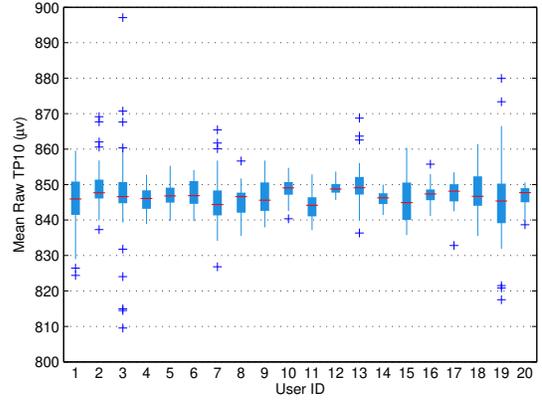}
\caption{User-wise distribution of a low significant feature}
\label{Fig:meanrawTP10} \vspace{-4mm}
\end{figure}

The two least important features; \emph{zero crossing rate (ZCR)} of \emph{Alpha} wave from TP9 and \emph{zero crossing rate (ZCR)} of \emph{raw EEG} from TP9 consist of mainly \emph{zeros}. We show the distribution of the third least important feature in Figure~\ref{Fig:meanrawTP10}. As the figure shows, for many users mean \emph{raw EEG} from TP10 behaves similar (i.e. less predictive power).



\begin{table}[h!]
\scriptsize
\centering
\caption{Top-20 features by importance} \vspace{-1mm}
\begin{tabular}{p{0.6cm}|p{2.2cm}|p{0.6cm}|p{2.3cm}}
{\bf Index } & {\bf Feature} & {\bf Index} & {\bf Feature}\\ \hline
1 & Mean Alpha TP9& 11& Min Beta TP9 \\ \hline
2 & Mean Alpha AF7 & 12 &Mean Gamma AF7 \\ \hline
3 & Mean Alpha TP10 & 13 &Max Alpha TP10  \\ \hline
4 & Mean Beta TP9 & 14 &Min Alpha TP10  \\ \hline
5 & Max Alpha TP9 & 15 & Min Raw TP10 \\ \hline
6 & Mean Theta AF8 & 16 & Mean Gamma TP10\\ \hline
7 & Mean Theta AF7 & 17 &Mean Gamma AF8\\ \hline
8 & Mean Alpha AF8 & 18 & Mean Beta AF7 \\ \hline
9& Max Raw TP10 & 19 & Mean Theta TP10 \\ \hline
10 & Mean Beta TP10 &20& Mean Beta AF8\\ \hline
\end{tabular} \vspace{-2mm}
\label{Tab:FeatureImportance}
\end{table}

\vspace{-4mm}
\subsection{Electrode placement}

\begin{figure}[h]
\centering \vspace{-4mm}
\subfloat[Same song]{\label{Fig:ElecPlacementSame}\includegraphics[trim=0cm 0cm 0cm 0cm, clip=true,scale=0.22]{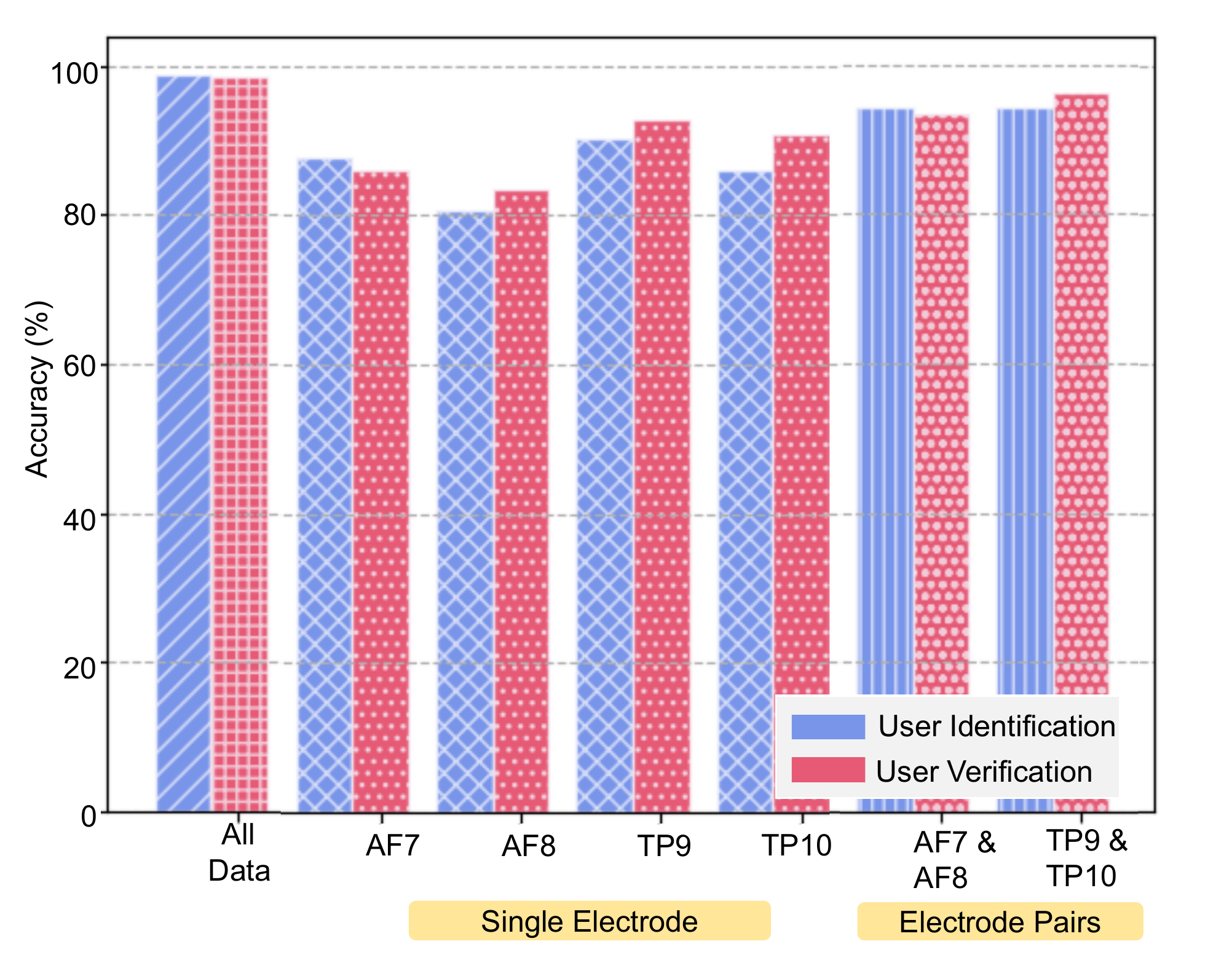}}
\subfloat[Favourite song]{\label{Fig:ElecPlacementFavorite}\includegraphics[trim=0cm 0cm 0cm 0cm, clip=true,scale=0.22]{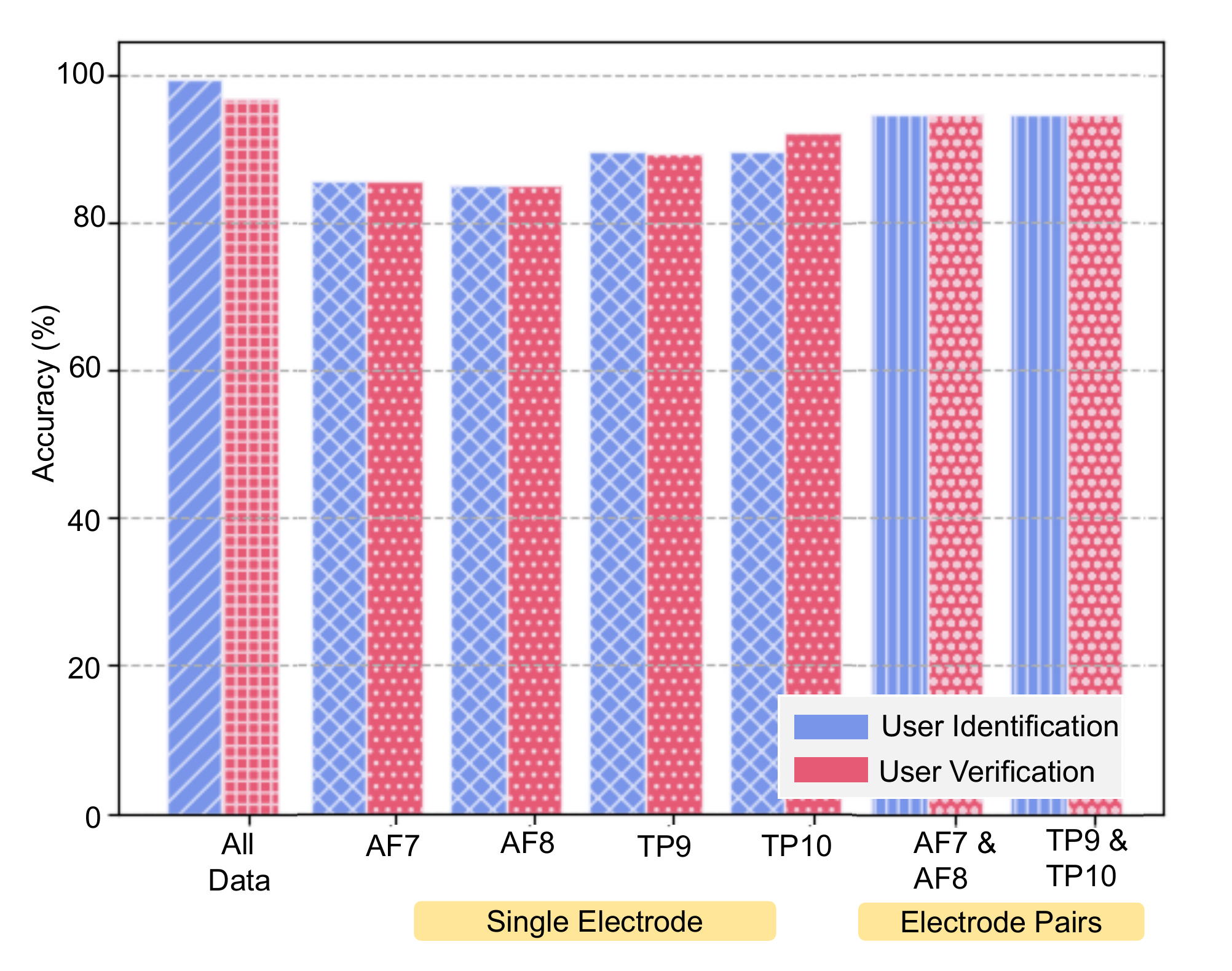}}
\caption{Accuracy using different electrodes} \vspace{-5mm}
\label{Fig:Electrodes}
\end{figure}

We next obtain the accuracy for using the data collected from individual electrodes and electrode combination scenarios of two front electrodes and two rear electrodes. We used selected electrodes' features from 80 feature vector of the total dataset given in Table~\ref{Tab:Samples}. In Figure~\ref{Fig:ElecPlacementSame} and Figure~\ref{Fig:ElecPlacementFavorite} we show the accuracies for the two experiment scenarios for both user verification and user identification. As expected when the number of electrodes decreases the accuracy drops and results show that there are no significant differences in accuracy between individual electrodes. However, combining two electrodes provides an accuracy of approximately 95\% which is only a 4\% difference to the all-electrode accuracy. This is a promising result as, it is easier to include two electrodes in the front of many consumer headsets. 


\vspace{-4mm}
\subsection{Significance of different brainwave frequencies}

\begin{figure}[h]
\centering \vspace{-4mm}
\subfloat[Same song]{\label{Fig:FrequencySame}\includegraphics[trim=0cm 0cm 0cm 0cm, clip=true,scale=0.22]{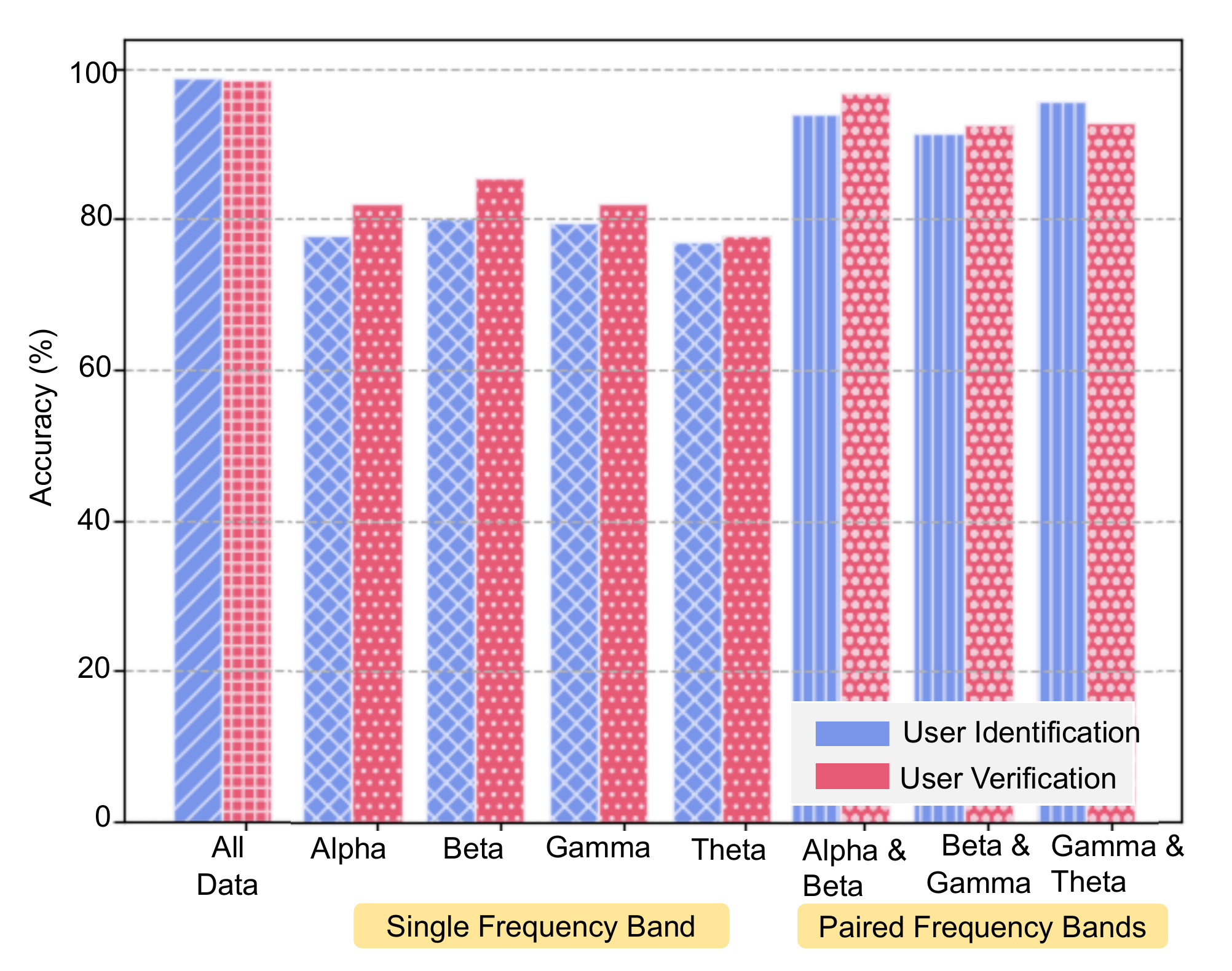}}
\subfloat[Favourite song]{\label{Fig:FrequencyFavorite}\includegraphics[trim=0cm 0cm 0cm 0cm, clip=true,scale=0.22]{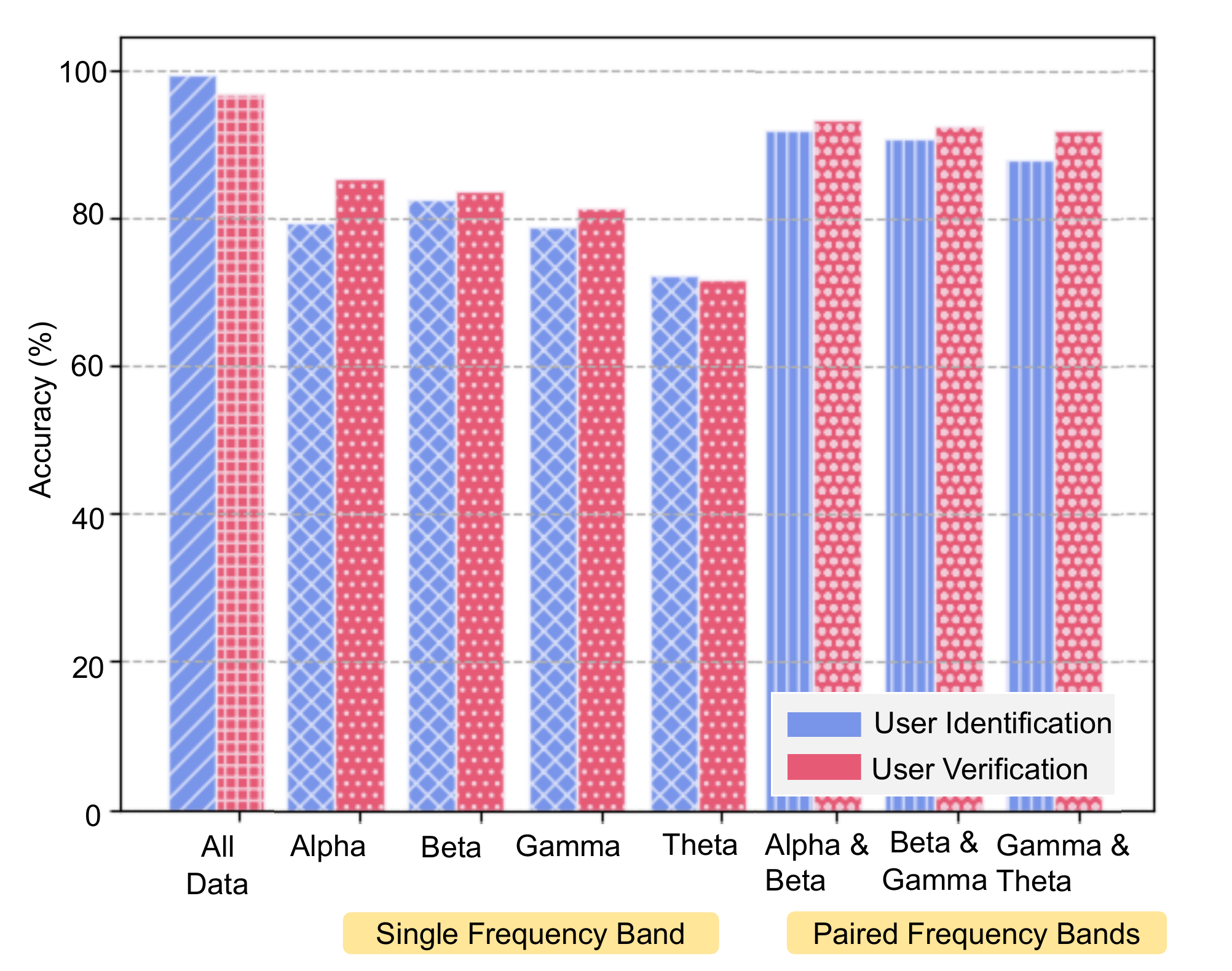}}
\caption{Accuracy using different brainwave frequencies}\vspace{-4mm}
\label{Fig:Frequencies}
\end{figure}

We next analyse whether some brainwaves show more differentiation between individual participants. We trained classifiers only from data from different brainwave types and some combinations of them and the accuracies are shown in Figure~\ref{Fig:Frequencies} for the two experiment scenarios. As the figures show none of the individual waves gives sufficient accuracy. Also, it is interesting to observe that despite not showing up into overall top-20 features ({\bf c.f} Section~\ref{SubSec:Importance}) \emph{Gamma} waves provide a performance in par with \emph{Alpha} and \emph{Beta} waves. Figures~\ref{Fig:FrequencySame} and \ref{Fig:FrequencyFavorite} show that combining \emph{Alpha} and \emph{Beta} waves provide a slightly higher performance compared to other combinations. This can be expected as there were more \emph{Alpha} and \emph{Beta} related features in top-20 significant features.


\subsection{Cross song and combined data training and testing}
\label{SubSec:Cross}

Our classifier results shown previously indicated that both the experiments; listening to same song and favorite song provide equally good results. To check whether there is actually a difference in brainwave responses in two scenarios, we did a cross experiment training and testing; i.e. training a model using data from listening to \emph{same song} and testing the model's performance on test data from listening to the \emph{favorite song} and vice versa. We show the results in Table~\ref{Tab:CrossSong}. The results show that there is approximately 20\%-30\% drop in accuracy in cross song setting. It is interesting to note that models are still achieve an accuracy in the range between 64\%-79\% indicating there is some common response to music in the brain.

\begin{table}[h]
\scriptsize
\centering
\caption{Cross song training and testing } \vspace{-1mm}
\begin{tabular}{p{1.6cm}|p{1.6cm}|p{1.6cm}}
& \multicolumn{2}{|c}{\textbf{Tested on}} \\ \hline
\multicolumn{3}{l}{\textbf{User identification}} \\ \hline 
{\bf Trained on} & \textbf{Favorite Song} &  \textbf{Same Song}  \\ \hline
Favorite Song &  99.46\% &  79.57\%\\ \hline
Same Song  & 72.04\% & 98.39\% \\ \hline
\multicolumn{3}{l}{\textbf{User verification}} \\ \hline 
Favorite Song & 97.31\%  & 77.96\% \\ \hline
Same Song  & 64.52\% &  98.92\%\\ \hline
\end{tabular} \vspace{-2mm}
\label{Tab:CrossSong}
\end{table}

Finally, we trained a classifier with the combined training dataset and tested on different test datasets. For the combined test dataset the classifier gave an accuracy of 98.39\% and individual accuracies of 98.92\% and 97.85\% respectively on same song and favorite song test datasets for user identification.

\subsection{Summary of results}

In Table~\ref{Tab:Summary} we provide a summarized view of all of our results and we below list the main take away messages.

\begin{itemize}
\item Individuals can be authenticated accurately using brainwaves extracted from audio stimulation of the brain and we were able to achieve accuracies in the range of 97\%-99\% for both user identification and verification tasks.

\item The accuracies drop only by 2\%-5\% when only two electrodes are used. This is promising as it is feasible to include two electrodes in the front side of many consumer headsets.

\item Feature importance analysis and frequency band-wise training indicated that features derived from \emph{Alpha} and \emph{Beta} waveforms have more predictive capability compared to other frequency bands.

\end{itemize}

\begin{table}[h]
\scriptsize
\centering
\caption{Summary of results } \vspace{-1mm}
\begin{tabular}{p{1.6cm}|p{1.3cm}|p{1.2cm}|p{1.3cm}|p{1.2cm}}
& \multicolumn{2}{|c|}{\textbf{Favorite song}} & \multicolumn{2}{|c}{\textbf{Same song}} \\ \hline
{\bf Scenario} & {\bf Identification} & {\bf Verification}  & {\bf Identification} & {\bf Verification}  \\ \hline
All data &  {\bf 99.46\% }&{\bf 97.31\%}& {\bf 98.39\%} & {\bf 98.92\%} \\ \hline
\multicolumn{5}{l}{\textbf{Electrode position}} \\ \hline 
AF7  & 86.02\% &86.02\% & 87.63\%&86.02\%\\ \hline
AF8 & 85.48\% & 85.48\% &80.10\% &  83.87\%\\ \hline
TP9 & 89.78\% & 89.24\% & 90.86\%& 92.01\%\\ \hline
TP10 & 89.78\% &92.47\%  &86.55\% &90.86\% \\ \hline
AF7 + AF8 & 94.62\% & 94.62\% &94.62\% &93.54\% \\ \hline
TP9 + TP10 & {\bf 94.62\%} & {\bf 94.62\%}   & {\bf 94.62\%}& {\bf 96.77\%}\\ \hline
\multicolumn{5}{l}{\textbf{Brainwave type}} \\ \hline 
Alpha & 79.56\% & 85.48\%  &  77.96\% & 82.26\%  \\ \hline
Beta &  82.80\%&83.87\%  & 80.11\% & 85.48\% \\ \hline
Gamma & 79.03\% & 81.72\%  & 79.57\% & 82.26\%  \\ \hline
Theta &  72.58\%&72.04\%  & 76.88\%& 77.96\%   \\ \hline
Alpha + Beta  &{\bf 91.93\% } &{\bf 93.54\%}&{\bf 94.09\%} & {\bf 96.77\%}\\ \hline
Beta + Gamma  &90.86\%  &92.47\%&91.40\% & 92.47\%\\ \hline
Gamma + Theta   & 88.17\% & 91.93\%&95.70\%& 93.01\%\\ \hline
\end{tabular} \vspace{-2mm}
\label{Tab:Summary}
\end{table}



\section{Related Work}
\label{Sec:RelatedWork}


\subsection{Novel behavioral biometric modalities}
Limitations of the passwords/token-based systems as well as static biometrics such as fingerprints led to a large body of works exploiting behavioral biometrics as a mean of user authentication. Such biometrics include gait~\cite{gafurov2006biometric},
touch patterns~\cite{frank2013touchalytics},
breathing acoustics~\cite{chauhan2017breathrnnet,chauhan2018performance}, and heart rate~\cite{article}. 
For example, Buriro et al.~\cite{BURIRO201989} proposed an authentication mechanism for smartphones using the built-in smartphone sensors to capture the behavioural signatures of the way a user slides the lock button on the screen to unlock the phone, and brings the phone towards ear. True Acceptance Rate of 99.35\% was achieved using Random Forest classifier with 85 users performing the unlocking actions, in sitting, standing, and walking postures. Hadiyoso et al.~\cite{article} demonstrated  a new biometric modality using ECG signals from 11 participants. Using empirical mode decomposition (EMD) and statistical analysis for feature extraction authors achieved an accuracy of 93.6\% using discriminant analysis.

More recently another set of modalities that extend the conventional \emph{``what you are''} definition of behavioral biometrics to \emph{``what you perceive''}, are becoming popular as they tend to be more resilient, reliable, and non-intrusive. Also, such biometrics suit well for IoT devices that have limited user interactivity. 

For instance, Loulakis et al.~\cite{loulakis2017quantum} proposed a \emph{quantum biometric} method using single photon detection ability of the human retina. Using photon counting principles of human rod vision, authors formed light flash patterns that only a specific individual can identify and theoretically showed that false positive and false negative rates are lower than $10^{-10}$ and $10^{-4}$. 
Finn et al.~\cite{finn2015functional} demonstrated that functional connectivity profiles of the brain observed by fMRI can potentially be unique to individuals. Using data collected from 126 participants, authors showed that between 87\%-99\% accuracy can be achieved under different brain activities.

Our work contributes to behavioral biometrics of \emph{``what you perceive''} category by exploring the feasibility of using music induced brainwave patterns as a novel authentication modality.





\vspace{-3mm}
\subsection{Use of EEG for user authentication}

Several work looked different forms of brain stimulations that can be used to generate EEG data for user authentication such as 
\emph{visual}
~\cite{sym10110537}, \emph{audio-visual}
{~\cite{s19071664}},
\emph{mental tasks}~\cite{chuang2013think},~\cite{bajwa2016neurokey} as well as \emph{resting state}~\cite{sohankar2015bias}.

Chiu et al.~\cite{sym10110537} used eight channel EEG headset (BR8) to capture brainwaves from 30 subjects. Using visual stimulated brainwave data from a set of 15 images (5 familiar, 5 déjà vu, and 5 unfamiliar) authentication tokens were retrieved and authors achieved an accuracy of 98.7\% using an SVM classifier. Huang et al.~\cite{s19071664} used 14 channel Emotive Epoc+ headset to focus on brainwaves evoked by an audio-visual paradigm of subject’s own face image and a voice calling their own name. With raw EEG collected from 30 adult subjects and a bagging-based ensemble machine learning model authors achieved an accuracy around 92\%. 




Chuang et al.~\cite{chuang2013think} used a single channel EEG headset (Neurosky) for capturing brainwaves from 15 subjects. Using 7 mental tasks in two repeated sessions, an accuracy of 99\% was achieved with a cosine distance based kNN classifier. 
Sohankar et al.~\cite{sohankar2015bias} also used Neurosky headset, but focused on brain behavior while the user is in resting state and achieved 95\% accuracy for user authentication and 80\% for user identification with 10 different subjects. Similarly, MindID~\cite{zhang2017mindid} exploited Delta brainwaves for authentication (98.2\% accuracy), while users relax with closed eyes using Emotive Epoc+ with 8 subjects. In another study using the same device with 12 subjects, Jayarathne et al.~\cite{jayarathne2016brainid} showed 96.67\% accuracy can be achieved using LDA for classification. Zeynali et al.~\cite{ZEYNALI2019261} conducted a similar study using five mental activities and achieved accuracies in the range of 97-98\% using Neural Network classifier. \textit{Our work is different from these related work as we introduce a new form of non-invasive music based brain stimulation method for EEG based user authentication.}

Similar to our work, Kaur et al.~\cite{kaur2016novel} explored the potential of using music stimulations for authentication. An experiment with 60 users listening to 4 genres of music, 97.5\% and 93.83\% of accuracies were reached using Hidden Markov Models and SVM classifiers respectively. Huang et al.~\cite{huang2019eeg} proposed a system targeting IoT applications. In contrast to this work, we use a different form of audio stimulation and our experiments span over 1.5 months with multiple sessions.



\vspace{-3mm}
\subsection{Authentication solutions for head-mounted devices}

Head-mounted devices of different types such as smart glasses, smart ear-buds, and virtual/augmented reality headsets are increasingly becoming standalone devices~\cite{sen2017}. Due to the lack of user input modalities, several work looked into the possibility of coming up with password-less authentication mechanisms for head-mounted devices.
Head movement patterns of individuals generated in response to an audio stimulus was proposed to be used to authenticate them to smart glasses by Li et al.~\cite{li2016whose}. With accelerometer data from 30 users and simple distance-based threshold classifiers, the authors showed Equal Error Rate of 4.43\% can be achieved.  
Using the data collected from 20 users, Rogers et al.~\cite{rogers} proposed blinks and head-movements captured using infrared, accelerometer, and gyroscope sensors can be used for user identification with an accuracy of 94\%. Mustafa et al.~\cite{mustafa} proposed security-sensitive Virtual Reality (VR) using head, hand and (or) body movement patterns exhibited by a user freely interacting with a VR. With 23 users the authors achieved a mean equal error rates of 7\%. George et al.~\cite{george2019investigating} investigated the third dimension for authentication in immersive virtual
reality and in the real world. In the proposed system, users’ authenticate by selecting a series of 3D objects in a room using a pointer and investigated the influence of randomized user and object positions, in a series of user studies with 48 subjects.

In contrast to above work, this paper propose EEG as a feasible authentication modality for head-mounted devices and to the best of our understanding this is the first work that demonstrated the feasibility of using a commodity brainwave headset for user authentication.



\section{Discussion \& Future Work}
\label{Sec:Discussion}

We below discuss the implications of our results, potential application scenarios, limitations, and potential future work. 

\textbf{Application scenarios.}
Our results showed that it is possible to authenticate users with over 97\% accuracy by capturing the brainwave patterns generated whilst a person is listening to music. The results are promising and the fact that we were able to do it using a commodity-headset allude that brainwaves can be used as a potential \emph{entry point} and \emph{continuous authentication} solution for smart-headsets. Also, we showed that two electrodes are sufficient to achieve around 95\% accuracy. This is significant as most of the commodity head-mounted devices such as smart glasses, ear-buds, and VR headsets can easily support two front electrodes. We believe the innovative applications that are expected to rise along with the IoT in education, remote support, tele-health, and many other domains will demand for increased security mechanisms for head-mounted devices. To this end, MusicID provides an intrusion-free and secure solution. 

\textbf{Participant pool.}
We used a dataset collected from 20 voluntary participants spanning over 1.5 months. 55\% of the participants provided data in over more than three sessions and 45\% provided data in only two sessions. While we achieved over 97\% accuracy, it is necessary to explore the feasibility of MusicID in a much larger participant pool that includes diverse demographics. However, we highlight that majority of comparable studies (E.g. \cite{chuang2013think, sohankar2015bias, zhang2017mindid, jayarathne2016brainid, li2016whose}) demonstrated their accuracies with less than 20 participants.

\textbf{Contextual changes.}
Our results showed consistency over multiple sessions spanning 1.5 months and we were able to re-identify users in new sessions based on previous session's data.  Nonetheless further experiments on collecting data after different physical and mental activities, different times of the day can help to identify whether there are context-driven short term variations that are dependent on physical activity, health conditions~\cite{roizenblatt2001alpha}, stress levels, fatigue~\cite{craig2012regional}, and mood. Also, there can be long term variations in brainwave responses (e.g. due to aging~\cite{anyanwu2007neurochemical}) and a more robust model must be tuned over time to capture such variations. For such settings learning models based on \emph{transfer learning} can be leveraged.
 
\textbf{Attacks.}
Brainwave patterns provide a unique biometric that is more secure compared to other biometrics. For example, it is difficult to record brainwave patterns for replay attacks compared to other biometrics. For example, a person's fingerprint can be copied while the person is asleep, however this type of recording is not applicable for brainwaves as brain activities are completely different whilst sleeping. Similarly, an attacker can record a person's voice from a distance and use voice synthesis to bypass a voice based authentication system. However, brainwaves can't be recorded at a distance and also has to be collected in the very specific moments where the user is engaged in a related activity such as listening to music in this case. One possible attack is an attacker starting with some existing data and trying to synthesize the target's brainwave pattern, which appears to be highly difficult.

\vspace{-3mm}
\subsection{Future work}

During our study we  found another potential biometric modality related to the electrical activities in eye and has
the potential to be used as a standalone modality or in combination with EEG. Due to the potential difference between the cornea and retina of the eye, eye blinks evoke an artifact waveform in EEG known as EOG ({\bf E}lectro{\bf o}culo{\bf g}ram) which can potentially be unique to individuals. The usage of EOG as a biometric system has been tested in~\cite{7506818} using electrodes placed on the face to capture the potential difference evoked due to eye blinks. During our study we found that Muse headset is capable of capturing the EEG signals during eye blinks, which can then be used to extract EOG using signal processing techniques such as Discrete Wavelet Transformation.   






\section{Conclusion}
\label{Sec:Conclusion}

We  proposed MusicID a behavioral biometric modality suitable for smart-headsets enabled IoT environments induced by the human brain's response to music. Using a 4-electrode commodity brainwave headset we collected brainwave data samples from real users over multiple sessions while they were listening two forms of music; \emph{i) a common English song}, \emph{ii) individual's favorite song}. We built Random Forest classifiers and showed that an accuracy over 98\% can be achieved for \emph{user identification} and an accuracy over 97\% can be achieved for \emph{user verification} using both audio stimulations. Our feature importance analysis showed that \emph{Alpha} and \emph{Beta} waves have more predictive capabilities. By investigating the classifier performance at individual electrodes and in combinations, we showed that use of two electrodes instead of four, drops accuracy only by 2\%-5\%. While further studies are required with much larger set of users, the performance of MusicID indicates the feasibility of a non-obtrusive, user friendly, and secure solution to the problem of \emph{entry point} and \emph{continuous} user authentication in smart-headsets that are expected to become increasingly popular with the proliferation of IoT.   

\vspace{-3mm}
\bibliographystyle{IEEEtran}
\bibliography{IEEEabrv,bibliography.bib}
\end{document}